\documentclass[12pt]{article}

\begin{document}

\begin{center}
 {\bf MARRIAGE OF ELECTROMAGNETISM AND GRAVITY

  IN EXTENDED SPACE MODEL

  AND ASTROPHYSICAL PHENOMENA}
\end{center}

\begin{center}
V.A. ANDREEV
\end{center}
{\it Lebedev Physical Institute of the Russian Academy of Sciences,\\
Moscow, Russia. E-mail: andrvlad@yandex.ru}

\begin{center}
D.Yu. TSIPENYUK
\end{center}
{\it Prokhorov General Physics Institute of the Russian Academy of Sciences,\\
Moscow, Russia.  E-mail: tsip@kapella.gpi.ru}

\begin{abstract}
{\small The generalization of Einstein's special theory of
relativity (SRT) is proposed. In this model the possibility of
unification of scalar gravity and electromagnetism into a single
unified field is considered. Formally, the generalization of the
SRT is that instead of (1+3)-dimensional Minkowski space the
(1+4)-dimensional extension G is considered. As a fifth additional
coordinate the interval S is used. This value is saved under the
usual Lorentz transformations in Minkowski space M, but it changes
when the transformations in the extended space G are used. We call
this model the extended space model (ESM). From a physical point
of view our expansion means that processes in which the rest mass
of the particles changes are acceptable now. If the rest mass of a
particle does not change and the physical quantities do not depend
on an additional variable S, then the electromagnetic and
gravitational fields exist independently of each other. But if the
rest mass is variable and there is a dependence on S, then these
two fields are combined into a single unified field. In the
extended space model a photon can have a nonzero mass and this
mass can be either positive or negative. The gravitational effects
such as the speed of escape, gravitational red shift and
deflection of light can be analyzed in the frame of the extended
space model.
 In this model all these gravitational effects can be found algebraically by
 the rotations in the (1+4) dimensional space.
Now it becomes possible to predict some future results of visible
size of supermassive objects in our Universe due to new stage of
experimental astronomy development in the RadioAstron Project and
analyze phenomena is an explosion of the star V838 Mon.}

\end{abstract}

{\it Keywords}: gravity, electromagnetism, interval, gravstar,
multy-dimensional space, star V838 Mon.

\section{Introduction}\label{aba:sec1}
We consider a generalization of Einstein's special theory of
relativity (STR) in a 5-dimensional space, or more specifically in
a (1+4)-dimensional space with a metric  (+ - - - -). However, it
is well known that the photon can be considered as a massless
particle, and  described by the plane wave only in an infinite
empty space ~\cite{LL1,SW2}.

But if a photon falls into the environment or  in confined space,
such as a resonator or waveguide, it acquires a nonzero mass, see
\cite{GN3,RV4}.

  Under the particle mass $ m $, we will understand its rest mass, which is a
   Lorentz scalar.
No other masses will not appear in this work. Here we follow the
 recommendations of \cite{OK5}.
Similarly, we can consider the process of changing the mass of
other particles, such as electrons, assuming that it depends on
external conditions and influences.

 Thus, it seems natural to expand the space of parameters characterizing
particle, taking into account  the fact that the interaction of
its mass can vary. We call it the extended space.

\section{The structure of the extended space}
Such particle having a mass $ m $, corresponds to a hyperboloid in Minkowski
space, in the limiting case this hyperboloid degenerates into a cone.
\begin{eqnarray}
\label{AL1}
  s^2\;=\;(ct)^2\;-\;x^2\;-\;y^2\;-\;z^2\;.
\end{eqnarray}

Since the change of the mass of a particle corresponds  its
transition from one yperboloid to the other, i.e. change of the
corresponding interval, it seems natural to choose interval  $ s $
as an additional fifth coordinate. Thus, we will work in a space
with coordinates
 $( t,x,y,z,s )$ and metric $(+ - - - - )$.
 The objects under consideration are located on a cone
\begin{eqnarray}
\label{AL2}
  (ct)^2\;-\;x^2\;-\;y^2\;-\;z^2\;-\;s^2\;=\;0\;.
\end{eqnarray}

We denote this space as $G(1,4)$. The Minkowski $M(1,3)$ space is a subspace
of $G(1,4)$. An interval in the Minkowski $M(1,3)$ space plays role of the
fifth coordinate in the $G(1,4)$ space. We designate this coordinate by the
letter $S$. The other coordinates are designated as $T,X,Y,Z$. One of the
characteristic features of this Extended Space Model (ESM) is that the
particle's rest mass $m_0$ is s variable quantity and a photon, moving in a
medium with refraction index $n>1$, acquires a nonzero mass. This mass can be
both positive and negative.

 The usual  (1+2)-dimensional cones and hyperboloids occur as sections of the
 surface (\ref{AL2})  by hyperplanes $ s = s_0$.
In the space $ G (1,4) $ one can constructed in usual way the
objects that have different tensor nature and transform
appropriately under linear transformations of the $G(1,4)$ space
\cite{KN6}.

In Minkowski space $ M(1,3) $ a 4-vector of energy and momentum
\begin{eqnarray}
\label{AL3}
 \tilde p\;=\;\left(\frac Ec,p_x,p_y,p_z\right)
\end{eqnarray}
is associated to each particle, \cite{LL1}.

In the extended space of $ G(1,4)$, we completes its to 5-vector
\begin{eqnarray}
\label{AL4}
\bar p\;=\;\left(\frac Ec,p_x,p_y,p_z,mc\right).
\end{eqnarray}
For free particles, the components of the vector (\ref{AL4}) satisfy the
equation
\begin{eqnarray}
\label{AL5}
 E^2\;=\;c^2p^2_x\;+c^2p^2_y\;+\;c^2p^2_z\;+\;m^2c^4.
\end{eqnarray}

It is well-known relation of relativistic mechanics, which relates
the energy, momentum and mass of a particle. Its geometric meaning
is that the vector (\ref{AL4}) is isotropic, i.e. its length in
the space $ G (1,4) $ is equal to zero. However, in contrary to
the usual relativistic mechanics, we now suppose that the mass  $
m $ is also a variable, and it can vary at motion of a particle on
the cones (\ref{AL2}),(\ref{AL5}). It should be understood so that
the mass of the particle changes when it enters the region of the
space that has a nonzero density of matter. Since in such areas
the speed of light is reduced, they can be characterized  by value
$ n $ - optical density. The parameter $ n $ relates the speed of
light in vacuum $ c $ with the speed of light in a medium $ v $.
\begin{eqnarray}
\label{AL18}
 v=\frac cn.
\end{eqnarray}

A set of variables (\ref{AL4}) forms a 5-pulse, its components are conserved,
if the space  $ G(1,4) $ is invariant under the corresponding direction.
In particular, its fifth component  $ p_4 $, having  sense of mass, does not
change if the particle moves  in the area with constant value  $ n $.

\section{The vectors of the free particles}
In the usual relativistic
mechanics and field theory the mass of a particle is constant, and for
particles with
zero masses and nonzero rest masses different methods of description are used.
The particles with nonzero rest masses are characterized by their mass $ m $
and speed $ \vec v $.
The particles with zero mass (photons) are characterized by frequency
$ \omega $ and wavelength $ \lambda $.
These $\omega$ and $\lambda$ are connected with energy $E$ and momentum
$\vec p$ sa follows
\begin{eqnarray}
\label{AL6}
E\;=\;\hbar\omega\;,\;\;\;\vec p\;=\;\frac{2\pi \hbar}{\lambda}\vec k.
\end{eqnarray}

The 4-vector
\begin{eqnarray}
\label{AL7}
 \tilde p=\left(\frac Ec,\vec p \right)=\left(\frac{mc}{\sqrt{1-\beta^2}},
 \frac{m\vec v}{\sqrt{1-\beta^2}}\right),\\
\nonumber
\beta^2=\frac{v^2}{c^2}.
\end{eqnarray}
corresponds to a particles with nonzero rest mass.

The 4-vector
\begin{eqnarray}
\label{AL8}
\tilde p=\left(\frac{\hbar\omega}c,\frac{2\pi \hbar}{\lambda}\vec k \right)=
\left(\frac{\hbar\omega}c,\frac{\hbar\omega}c\vec k\right).
\end{eqnarray}
corresponds to a particles with zero mass.

In the frame of our approach, there is no difference between
massive and massless particles, and therefore one can establish a
connection between these two methods of description. This can be
done using the relation (\ref{AL6}) and the hypothesis of de
Broglie, according to which these relations hold for the massive
particles. Now, substituting (\ref{AL6}) in (\ref{AL4}), we obtain
the relation between the mass $ m $, frequency $ \omega $ and
wavelength $ \lambda $

\begin{eqnarray}
\label{AL9}
 \omega^2\;=\;\left(\frac{2\pi c}{\lambda}\right)^2\;+\;\frac{m^2c^4}{{\hbar}^2}.
     \end{eqnarray}

\begin{eqnarray}
\label{AL10}
 \omega\;=\;\frac{mc^2}{\hbar\sqrt{1-\beta^2}} \;,\;\;
\lambda\;=\;\frac{2\pi \hbar}{mv} \sqrt{1-\beta^2}.
\end{eqnarray}
It follows that if $ v\to0\;\;\;\lambda\to\infty $, but
$ \omega\to\omega_0\neq0$.
Here  $ \omega_0$ determines the energy of a particle at rest.

Now we construct  5-vectors from  4-vectors (\ref{AL7},
\ref{AL8}).
 We suppose that a 5-vector
\begin{eqnarray}
\label{AL11}
 \bar p\;=\; (\;mc,\;\vec 0,\;mc\;)
 \end{eqnarray}
corresponds to a stationary particle of mass $ m $.

The 5-vector of a particle, which moves with velocity $ \vec v $, can be
obtained by transformation to the
moving coordinate system. Then the vector (\ref{AL11}) takes the form

\begin{eqnarray}
\label{AL12}
  \bar p\;=\;\left(\frac{mc}{\sqrt{1-\beta^2}}\;,\frac{m\vec v}
  {\sqrt{1-\beta^2}}\;,\; mc\;\right).
  \end{eqnarray}

Similarly the 4-vector (\ref{AL8}) transforms into 5-vector
\begin{eqnarray}
\label{AL13}
\bar p\;=\;\left(\frac{\hbar\omega}c\;,\;\frac{2\pi \hbar}{\lambda}\vec k \;,
0\; \right).
\end{eqnarray}

At the transition to a moving coordinate system the vector (\ref{AL13}) does
not change its form, only the frequency  $\omega$ changes its value.
\begin{eqnarray}
\label{AL14}
\omega\;\to\; \omega'= \frac\omega{\sqrt{1-\beta^2}}.
 \end{eqnarray}

Thus, in empty space in a stationary reference frame there are two
fundamentally different object with zero and nonzero masses, which in the
space of $ G (1,4) $ correspond to the 5-vectors

\begin{eqnarray}
\label{AL15}
  \left(\;\frac{\hbar\omega}c\;,\;\frac{\hbar\omega}c\;,\;0\;\right)
 \end{eqnarray}
and
\begin{eqnarray}
\label{AL16}
 \left(\;mc\;,\;0\;,\;mc\;\right).
\end{eqnarray}

For simplicity, we write  the vectors (\ref{AL15}), (\ref{AL16})
in (1+2)-dimensional space.  The vector (\ref{AL15}) describes a
photon with zero mass, the energy $ \hbar \omega $, and the
velocity  $ c $. The vector (\ref{AL16}) describes a stationary
particle of mass $ m $. The photon has a momentum $ p = \frac
{\hbar \omega} c $, a massive particle has a momentum equal to
zero. In the 5-dimensional space, these two vectors are isotropic,
in Minkowski space only the vector (\ref{AL15}) is isotropic.

The length of the vector $(x_0,x_1,x_2,x_3,x_4)$ is equal to
$$ l^2\;=\; x_0^2-x_1^2-x_2^2-x_3^2-x_4^2  $$

If we restrict ourselves to
Lorentz transformations in Minkowski space it is impossible to transform an
isotropic vector into anisotropic one and vice versa. In other words in frame
of the SRT photon can not acquires mass, and a massive particle can not be a
photon. But in the extended space $ G (1,4) $ a photon and a massive particle
can be related to each other by a simple rotation.

Asit  was already mentioned the parameter $n$ connects the speed
of light in vacuum with that in the medium: $v=c/n.$ Using it, one
can parametrize the fifth coordinate in the $G(1,4)$ space. The
value $n=1$ corresponds to the empty Minkowski space $M(1,3)$ in
which light moves at the velocity $c$. The propagation of light in
a medium with $n\neq 1$ is interpreted as an exit of a photon from
the Minkowski space and its transition into another subspace of
$G(1,4)$ space. This transition can be described as a rotation in
the $G(1,4)$ space. All types of such rotations were studied in
\cite{TA7}.

For hyperbolic rotation through the angle $\theta$ in the $(TS)$
plane the photon 5-vector (\ref{AL15}) with zero mass is
transformed in the following manner \cite{TA7}:
\begin{eqnarray}
\label{AL515}
  \left(\frac{\hbar\omega}c,\frac{\hbar\omega}c,0\right)\;\Rightarrow\\
  \nonumber
   \left(\frac{\hbar\omega}c\cosh\theta,\frac{\hbar\omega}c,\frac
   {\hbar\omega}c\sinh\theta\right)=\\
   \nonumber
 \left(\frac{\hbar\omega}cn,\frac{\hbar\omega}c,\frac{\hbar\omega}c
 \sqrt{n^2-1}\right).
 \end{eqnarray}
As a result of this transformation a particle with mass
\begin{eqnarray}
\label{AL516}
m=\frac{\hbar\omega}{c^2}\sinh\theta=\frac{\hbar\omega}{c^2}\sqrt{n^2-1}
   \end{eqnarray}
arise.
The velocity of this particle is defined by formula (\ref{AL18}).

Under the same rotation the massive 5-vector (\ref{AL16}) is transformed as
\begin{eqnarray}
\label{AL517}
\left(mc,0,,mc\right)\;\Rightarrow \left(mce^{\theta_{\pm}},0,
mce^{\theta_{\pm}}\right);\\
\nonumber
e^{\theta_{\pm}}=n\pm\sqrt{n^2-1}.
\end{eqnarray}
Under such rotation a massive particle changes its mass
\begin{eqnarray}
\label{AL518}
m \rightarrow me^{\theta},\;\;0\leq \theta < \infty
\end{eqnarray}
and energy but conserves its momentum.

Upon rotation through the angle $\phi$ in the (XS) plane the
photon vector is transformed in accordance to the law
\begin{eqnarray}
\label{AL519}
  \left(\frac{\hbar\omega}c,\frac{\hbar\omega}c,0,\right)\;\Rightarrow\\
  \nonumber
   \left(\frac{\hbar\omega}c\cos\phi,\frac{\hbar\omega}c,\frac{\hbar\omega}
   c\sin\phi\right)=\\
   \nonumber
 \left(\frac{\hbar\omega}c,\frac{\hbar\omega}{cn},\frac{\hbar\omega}{cn}
 \sqrt{n^2-1}\right).
 \end{eqnarray}
Given this, the photon acquires the mass
\begin{eqnarray}
\label{AL520}
m = \frac{\hbar\omega}{c^2}\sin\phi=\frac{\hbar\omega}{c^2n},
\end{eqnarray}
and velocity
\begin{eqnarray}
\label{AL521}
v=c\cos\phi=\frac cn.
\end{eqnarray}
The vector of a massive particle is transformed in accordance to  the law
\begin{eqnarray}
\label{AL522}
\nonumber
\left(mc,0,mc\right)\rightarrow\left(mc,-mc\sin\phi,mc\cos\phi\right)=\\
 \left(mc,-\frac{mc}n\sqrt{n^2-1},\frac{mc}n\right).\;\;\;\;\;
\end{eqnarray}
In this transformation the energy of a particle is conserved but its mass and
momentum change
\begin{eqnarray}
\label{AL523}
m \rightarrow m\cos\phi=\frac mn,
\end{eqnarray}
\begin{eqnarray}
\label{AL524}
0\rightarrow -mc\sin\phi=-\frac{mc}n\sqrt{n^2-1}.
\end{eqnarray}

The important fact is that the photon mass generated by
transformations (\ref{AL515},\ref{AL519}) can have either positive
and negative sign. This immediately follows from the symmetry
properties of $G(1,4)$ space. As to the particles that initially
had positive mass, after transformations (\ref{AL517},\ref{AL522})
it remains positive.

\section{Electrodynamics and gravitation in the extended space}
The source of the electromagnetic field is a current. In the
traditional formulation of the electromagnetic theory the current
is described by a 4-vector in Minkowski space $M(1,3)$ \cite{LL1}.
\begin{eqnarray}
\label{AL17}
 \tilde \rho=\left(\rho,\; \vec j \right)=\left(\frac{\rho_0 c}
 {\sqrt{1-\beta^2}},
\frac{\rho_0\vec v}{\sqrt{1-\beta^2}}\;\right),\\
\nonumber
\beta^2=\frac{v^2}{c^2},\;{\tilde \rho}^2=c^2\rho_0^2.
\end{eqnarray}

Here $\rho_0(t,x,y,z)$ - is an electric charge density in the point
$(t,x,y,z)$
in the space  $M(1,3)$, and $(v_x(t,x,y,z),v_y(t,x,y,z),v_z(t,x,y,z)$ - is a
local velocity of a charge density.

At the transition to the extended $G(1,4)$ space it is necessary to change
a (1+3)-current vector $ \tilde \rho$ by a (1+4)-vector $ \bar \rho$.
In accordance with the principles of the developed model, an additional c
oordinate of the vector $ \bar \rho $ must be an isotropic (1+4)-vector.
In addition, we want our model describes both the electromagnetic and
gravitational field, so the fifth component of the current should be defined
so that it be the source of the gravitational field.

We suppose  that the source of a unit electromagnetic and gravitational field,
is a particle which has both a mass and a  charge. In this case, we assume
that the mass may not have any charge, but the charge should always have a mass. In our model we assume that the charge is constant and does not change under transformations of the rotation group $ L(1,4)$ of the extended space $ G (1,4) $. And the rest mass, which is a scalar with respect to the Lorentz group, is a fifth component of the vector with respect to the group of $L(1,4) $.

We want to construct a 5-dimensional current vector $\bar\rho$ as a
generalization of  4-dimensional current vector $ \tilde\rho $. To do this one
must add one component to it.

In the ordinary electrodynamics 4-dimensional current vector
$\tilde \rho $ has the form (\ref{AL17}). Its structure it similar
to structure of the energy-momentum vector (\ref{AL7}) of the
particle, having a rest mass. The difference between them is that
in the vector (\ref{AL17}) instead of the rest mass $ m_0 $ there
is a local density of charge $\rho_0$. In the  extended space  $
G(1,4) $ we consider the energy-momentum-mass vector (\ref{AL17})
instead  of the energy-momentum vector (\ref{AL7}).

Thus, the 5-dimensional current vector, generating a unit
electro-gravitational
field, has the form
\begin{eqnarray}
\label{AL19}
\bar \rho=\left(j_0, \vec j,j_4 \right)=\left(\frac{emc}{\sqrt{1-\beta^2}},
\frac{em\vec v}
{\sqrt{1-\beta^2}},em c\right).
                               \end{eqnarray}
It is an isotropic vector
$$ {\bar \rho}^2=0.$$

The continuity equation, as in the usual case, is expressed by the
vanishing of the 5-divergence of the 5-current \ref{AL18}
\begin{eqnarray}
\label{AL20}
 \sum_{i=0}^4\frac{\partial j_i}{\partial x_i}=0.
                      \end{eqnarray}

If the charge is at rest the continuity equation takes the form
\begin{eqnarray}
\label{AL21}
 \frac{\partial m}{\partial t}+\frac{\partial m}{\partial x_4}=0.
\end{eqnarray}
Equation (\ref{AL21}) can be interpreted as the variation of the rest mass of
the particle by changing the properties of the environment.

In ordinary electrodynamics the law of conservation of charge follows from the
continuity equation
\begin{eqnarray}
\label{AL22}
 \frac{\partial }{\partial t}\int j_0dV=-\int\vec j d\vec n.
                          \end{eqnarray}
There is an integral over the volume in the left side of this equation, and
the right side - is the integral over the surface bounding this volume.

In electro-gravidynamics there is a  law of conservation of the value, which
is the product of the charge at the mass of a particle, which hold this charge. This law reads
\begin{eqnarray}
\label{AL23}
 \frac{\partial }{\partial t}\int j_0dV
=-\int\vec j d\vec n-\int\frac{\partial }{\partial x_4} j_4dV.
                          \end{eqnarray}
In this case, the change in the product of $em$ of a charge at the mass inside
a volume is defined as a stream of charged particles across the surface of the
volume and change of masses of  particles within the volume due to their
dependence on the coordinate $ x_4 $.

The current (\ref{AL19}) generates the electro-gravitational field
in the extended space of $ G (1,4) $. The potentials of this field
are determined by the equations \cite{TA7,TA8}.
\begin{eqnarray}
\label{AL24}
\Delta^{(5)} A_0\;=\;-4\pi\rho,
\end{eqnarray}

\begin{eqnarray}
\label{AL25}
\Delta^{(5)} \vec A\;=\;-\frac{4\pi}c \vec j,
\end{eqnarray}

\begin{eqnarray}
\label{AL26}
\Delta^{(5)}  A_s\;=\;-\frac{4\pi}c j_s.
\end{eqnarray}

Here
\begin{eqnarray}
\label{AL27}
\Delta^{(5)}\;=\;\frac{\partial^2}{\partial s^2}+\frac{\partial^2}
{\partial x^2}+
\frac{\partial^2}{\partial y^2}+
\frac{\partial^2}{\partial z^2}-\frac1{c^2}\frac{\partial^2}{\partial t^2}.
                                    \end{eqnarray}

With the help of the potentials $(A_0,A_x,A_y,A_z,A_s)$ one can
construct the tension tensor
\begin{eqnarray}
\label{AL28}
F_{ik}\;=\;\frac{\partial A_i}{\partial x_k}\;-
\;\frac{\partial A_k}{\partial x_i}\;;\;\;\;i,k=0,1,2,3,4.
                                      \end{eqnarray}

\begin{eqnarray}
\label{AL29}
\begin{array}{c}
\\
\\
||F_{ik}||\;=\;\\
\\
\\
\end{array}
\left(\begin{array}{ccccc}
0& -E_x& -E_y& -E_z& -Q\\
E_x& 0& -H_z& H_y& -G_x\\
E_y& H_Z& 0& -H_x& -G_y\\
E_z& -H_y& H_x& 0& -G_z\\
Q& G_x& G_y& G_z& 0
\end{array}\right)
 \end{eqnarray}
Here
\begin{eqnarray}
\label{AL30} Q=F_{40}=\frac{\partial A_4}{\partial
x_0}-\frac{\partial A_0}{\partial x_4}= \frac{\partial
A_s}{c\partial t}-\frac{\partial \varphi}{\partial s}.
 \end{eqnarray}
\begin{eqnarray}
\label{AL31}
G_x=F_{41}=\frac{\partial A_4}{\partial x_1}-\frac{\partial A_1}{\partial x_4}=
\frac{\partial A_s}{\partial x}-\frac{\partial A_x}{\partial s},\\
\nonumber
G_y=F_{42}=\frac{\partial A_4}{\partial x_2}-\frac{\partial A_2}{\partial x_4}=
\frac{\partial A_s}{\partial y}-\frac{\partial A_y}{\partial s},\\
\nonumber
G_z=F_{43}=\frac{\partial A_4}{\partial x_3}-\frac{\partial A_3}{\partial x_4}=
\frac{\partial A_s}{\partial z}-\frac{\partial A_z}{\partial s}.
\end{eqnarray}

Here is the equation satisfied by the intensity of $ F_ {ik} $. We'll call
them the extended Maxwell system.
The usual system of Maxwell equations consists of two pairs of equations,
which have fundamentally different structures. They are usually well known as
the first and second pair of Maxwell's equations. Extended system of Maxwell's
equations also consists of two types of equations fundamentally different
structure. We shall call them the equations of the first and second types.

The equations of the first type are formal consequence of the
formula (\ref{AL28}), which expresses the tension via potentials.
It follows immediately from their form that for any three indices
$ (i, j, k) $ the relation

\begin{eqnarray}
\label{AL32}
\frac{\partial F_{ij}}{\partial x_k}+\frac{\partial F_{ki}}{\partial x_j}+
\frac{\partial F_{jk}}{\partial x_i}=0
\end{eqnarray}
 is satisfied.

The validity of (\ref{AL32}) can be verified by direct
substitution  of the expression (\ref{AL28}) in the equation
(\ref{AL32}).
\begin{eqnarray}
\nonumber
\frac{\partial^2 A_i}{\partial x_k\partial x_j}\;-
\;\frac{\partial^2 A_j}{\partial x_k\partial x_i}\;+\;
\frac{\partial^2 A_k}{\partial x_j\partial x_i}\;-\\
\nonumber
\;\frac{\partial^2 A_i}{\partial x_j\partial x_k}\;+\;
\frac{\partial^2 A_j}{\partial x_i\partial x_k}\;-
\;\frac{\partial^2 A_k}{\partial x_i\partial x_j}\;=\;0.
\end{eqnarray}
There are exist 10 such equations. Let us consider now the
specific form of these equations, using the tension tensor
(\ref{AL29}).

If we restrict ourselves to  the sets of indices taking values
(0,1,2,3), then orresponding 4 equation are simply the first pair
of Maxwell's equations
\begin{eqnarray}
\label{AL33}
div \vec H=0, \;\;\;sim \;\;indices\;\; (1,2,3).
\end{eqnarray}

This is one equation. The three other equations which correspond to sets of
indices $ (0,1,2) $, $ (0,1,3), (0,2,3) $ form a unit vector equation
\begin{eqnarray}
\label{AL34}
rot\vec E\;+\;\frac1c\frac{\partial \vec H}{\partial t}\;=\;0.
\end{eqnarray}
Thus, the first pair of Maxwell's equations retain its form. In
the extended space $ G (1,4) $ another 6 equations are added to
them. Three of them, that are  corresponding to the sets $ (1,2,4)
$ $ (1,3,4) $, $ (2,3,4) $, can be  combined into one vector
equation

\begin{eqnarray}
\label{AL35}
rot\vec G\;+\;\frac{\partial \vec H}{\partial s}\;=\;0.
\end{eqnarray}
The other three triples $ (0,1,4) $ $ (0,2,4) $ $ (0,3,4) $ give us the three
remaining equations of the first type. They also can be merged into a single
vector equation
\begin{eqnarray}
\label{AL36}
 \frac{\partial \vec E}{\partial s}\;+\;\frac1c\frac{\partial \vec G}
{\partial t}+grad Q=0.
\end{eqnarray}
Thus, the equation of the first type from the exptended system of
Maxwell's equations in the space $ G(1,4) $ have the form
(\ref{AL33})-(\ref{AL36}). These 10 equations can be combined into
three vector equations and one scalar equation. Note that the
vector operators $ \; div, \; rot, \; grad \; $,  that appear in
these equations  have the usual three-dimensional form.

Let's  turn now to construction of the Maxwell equations of the
second type. These equations are follow from the equations for the
potentials (\ref{AL24})-(\ref{AL26}). However, it is necessary
first to impose the Lorentz gauge condition, which must satisfy
potential (\ref{AL23}). In the space  $ G (1,4) $ it has the form
\begin{eqnarray}
\label{AL37}
\frac1c\frac{\partial A_0}{\partial t}+\frac{\partial A_x}{\partial x}+
\frac{\partial A_y}{\partial y}+\frac{\partial A_z}{\partial z}+\frac
{\partial A_s}{\partial s}=0.
\end{eqnarray}

The second type of Maxwell's equations from  the extended system reads
\begin{eqnarray}
\label{AL38}
\sum_{k=0}^4 \frac{\partial F_{ik}}{\partial x_k}=-\frac{4\pi}{c}j_i\;;
                  \;\;\;i=0,1,2,3,4.
\end{eqnarray}
Substituting the elements of the tension tensor (\ref{AL29})into
the equation (\ref{AL38})  and taking into account the Lorentz
gauge condition (\ref{AL37}), one can obtain five vector
equations.

\begin{eqnarray}
\label{AL39}
div \vec E+\frac{\partial Q}{\partial s}\;=\;4\pi\rho, \;\;\;(i=0)
 \end{eqnarray}
\begin{eqnarray}
\label{AL40}
rot \vec H-\frac{\partial \vec G}{\partial s}-\frac1c\frac{\partial \vec E}
{\partial t}=\frac{4\pi}{c}\vec j,\;\;\;(i=1,2,3)
 \end{eqnarray}
\begin{eqnarray}
\label{AL41}
div \vec G+\frac 1c\frac{\partial Q}{\partial t}=4\pi j_4,
\;\;\;(i=4).
 \end{eqnarray}
The tension  tensor (\ref{AL29}) contains, in addition to the
components that are analogous to the usual electric and magnetic
fields, some additional components that describe gravitational
field. More precisely, in the case when the components of the
5-current (\ref{AL18}) depend on the coordinate $ x_4 $, all
components of (\ref{AL29}) describe a single electro-gravitational
field. If the current does not depend on the coordinate $ x_4 $,
the system of equations (\ref{AL34}), (\ref{AL36}), (\ref{AL38}),
(\ref{AL41}) splits into two systems. One of them is the system of
Maxwell's equations and the other is a Laplace equation for the
scalar gravitational field.

Thus, according to our model, in an empty space the gravitational
and electromagnetic fields exist as two different fields, but in the region
where there are particles and fields they form a unit
electromagnetic-gravitational field.

\section{Refraction index of a gravitational field}

Let's now study a problem of refraction index of a gravitational
field. Let there is a point mass, which gravitational field is
described by the Schwarzchild solution. We assume, that the
gravitational radius $r_g$ is small and we will consider all
effects at distance $ r > r_g$. In the literature there are
considered two expressions for refraction index $n$, appropriate
to the Schwarzchild field. One of them, we shall name it $n_1$, is
used in papers of Okun' \cite{OK11,OK12} and looks like
\begin{eqnarray}
\label{GL11}
n_1(r)=(g_{00})^{-1}=(1-\frac{r_g}r)^{-1}=1+\frac{2\gamma M}{rc^2}.
\end{eqnarray}
It can be found in the supposition, that in a constant
gravitational field the frequency of a photon $\omega$ remains
constant, but the wavelength  $\lambda$ and speed $v$ are varied.
The other refraction index $n_2$ one can get from the formula of
an interval in a weak gravitational field \cite{LL1}.
\begin{eqnarray}
\label{GL12}
ds^2\;=\;(c^2+2\varphi)dt^2-d\vec r^2.
\end{eqnarray}
Here $\varphi$ - is a potential of gravitational field. Supposing that
$d\vec r=\vec v dt$ and $ds^2=0$, one can find a speed of photon in a
gravitational field
\begin{eqnarray}
\label{GL13}
v\;=\;c \left(1+\frac{2\varphi}{c^2}\right)^{1/2}\;
\approx\;c\left(1+\frac{\varphi}{c^2}\right).
 \end{eqnarray}
 It is necessary here to take into account the fact that a potential of a
 gravitational field $\varphi$ - is negative.
For a point mass M we have
\begin{eqnarray}
\label{GL14}
\varphi(r)\;=\;-\frac{\gamma M}r.
\end{eqnarray}
Substituting the expression (\ref{GL14}) in the formula
(\ref{GL13}), one gets
\begin{eqnarray}
\label{GL15}
v\;\approx\;c\left(1-\frac{\gamma M}{rc^2}\right).
 \end{eqnarray}
Collins obtained the same formula in another way \cite{CL13}. He
considered a particle of mass $m_0$, located indefinitely far from
a point source of a gravitational field of mass M. Such particle
has an energy $E_0=m_0c^2$. When moving on a distance $r$ from a
source of a field, particles energy increases up to size
\begin{eqnarray}
\nonumber
E=m_0c^2 + \frac{\gamma m_0M}r.
\end{eqnarray}
 Collins offered to interpret this change of energy as change of a rest mass
 in a gravitational field.
\begin{eqnarray}
\label{GL16}
m\;=\;m_0\left(1+\frac{\gamma M}{rc^2}\right).
 \end{eqnarray}

Then he used a conservation law of a momentum $mv=m_0v_0$  and received the
law of change of speed in a gravitational field
\begin{eqnarray}
\label{GL17}
v\;=\;v_0\left(1+\frac{\gamma M}{rc^2}\right)^{-1}.
\end{eqnarray}

Supposing, that this law is valid also for photons, we get the formula for
change of photons speed in a gravitational field
\begin{eqnarray}
\label{GL18}
v\;=\;c\left(1+\frac{\gamma M}{rc^2}\right)^{-1}
\;\approx\;c\left(1-\frac{\gamma M}{rc^2}\right).
\end{eqnarray}

It is possible to interpret the formulas (\ref{GL15}), (\ref{GL18}) as hit of
a photon in medium with a refraction index
\begin{eqnarray}
\label{GL19}
n_2(r)\;=\;1+\frac{\gamma M}{rc^2}.
\end{eqnarray}

In the case, when the speed of a particle $ v$ is comparable with
the speed of light $c$, it is necessary to take into account in
the formula (\ref{GL16})  relativistic correction to a rest-mass
$m$ and to record it as
\begin{eqnarray}
\label{GL20}
M\;=\;m_0\left(1+\frac{\gamma M}{rc^2}+\frac{v^2}{2c^2}\right).
  \end{eqnarray}

Appropriate refraction index will look like
\begin{eqnarray}
\label{GL21}
n'\;=\;1+\frac{\gamma M}{rc^2}+\frac{v^2}{2c^2}.
 \end{eqnarray}

Such difference in definition of refraction index of a gravitational field is
connected with that the speech in these
cases goes about different objects, which differently interact with a
gravitational field. In ESM different rotations in extended space correspond
to these situations.

\section{Gravitational effects in ESM}

1) Speed of escape.
The speed of escape $v_2$ is that speed, which should be given to a body
located on a surface of the Earth, that it could
be deleted from Earth on an indefinitely large distance. Let $M$ - mass of the
Earth,$ $m - mass of a body located at the Earth surface, and $R$ - radius of
this surface. The expression for the speed of escape is
\begin{eqnarray}
\label{GL22}
v_2\;=\;\sqrt{2gR}\;=\;\sqrt{\frac {2\gamma M}R}.
\end{eqnarray}
We will receive now formula (\ref{GL22}) using ESM methods. Let's
consider a massive particle at rest, which removed to infinite
large distance from the Earth. Within the framework of our model
such particle is described by isotropic 5-vector of
energy-momentum-mass (\ref{AL4}). Space motion in gravitational
field along an axis X can compare movement in extended space G(1,
4) in a plane XS from a point with refraction index $ n = 1$ to
point with refraction index $n(r)$. Such motion is described by
some rotation in  expanded space  $ G(1,4) $. The rotation angle
express through refraction index $n$. Thus the massive particle at
rest acquires speed
$$
v=c\frac{\sqrt{n^2-1}}n.
$$
As to in this case we consider a motion of a massive body, so we assume
natural to use the refraction index $n_2$. Assuming, that it is close to unit,
i.e. that
\begin{eqnarray}
\label{GL23}
1>>\epsilon =\frac{\gamma M}{rc^2}
\end{eqnarray}
one can get, that
\begin{eqnarray}
\label{GL24}
V \approx  c\sqrt{2\epsilon }.
\end{eqnarray}
In case, when r = R - is the radius of the Earth, the formula
(\ref{GL24}) coincides with the formula (\ref{GL22}) and gives the
speed of escape $v = v_2$.

\bigskip 2) Red shift.

Gravitational red shift usually considered as a change of
frequency of a photon in the case of changing of a gravitational
field, in which photon is merged. In particular, at decreasing of
strength of a field the frequency of a photon also decreases, see
\cite{LL1}. However Okun' offered to recognize that not frequency
varies but wavelength of a photon varies, and just it named as red
displacement \cite{OK11,OK12}. Under our judgment both cases are
possible, but they correspond to different physical situations and
are described by different rotation angles express through
refraction index $n$. In the general theory of relativity the
formula that describes change of light frequency is, \cite{LL1}
\begin{eqnarray}
\label{GL25}
 \omega\;=\;\frac{\omega_0}{\sqrt{g_{00}}}\;\approx\;
\omega_0\left(1+\frac{\gamma M}{rc^2}\right).
 \end{eqnarray}
Here $\omega_0$ - is a frequency of a photon measured in universal time, it
remains constant at propagation of a beam of light.
And $\omega$ - is a frequency of the same photon which measured in its own
time. This frequency is various in various points of space. If the photon was
 emitted by a massive star, near to a star at small  $r$ the frequency of a
 photon is more, than far from it at large $r$. On infinity in the flat space,
 where there is no gravitational field, the universal time coincides with own
 and $\omega_0$  there is an observable frequency of a photon.

Let's consider now the same problem from the ESM point of view.
Within the framework of our model the isotropic 5-vector
(\ref{AL4}) is compared to a photon located in empty space.
Process of its movement to the point with refraction index $n$, at
which the change of frequency and energy happens, is described by
a rotation in (TS)  plane. At these rotations the photon vector
are transformed as
\begin{eqnarray}
\label{GL26}
\frac{\hbar\omega}c\left(1,1,0\right)\;\; \to\;
\frac{\hbar\omega}c\left(\cosh\theta,1,\sinh\theta\right)\;=\\
\nonumber
\frac{\hbar\omega}c\left(n,1,\sqrt{n^2-1}\;\right).
 \end{eqnarray}

One can see from here that the frequency $\omega_0$ of a photon in vacuum
and its frequency $\omega$ in a field, are connected by a ratio
\begin{eqnarray}
\label{GL27}
\omega\;=\;\omega_0\cosh\theta\;=\;\omega_0 n.
 \end{eqnarray}

We assume, that at calculation of change of photon frequency it is necessary
to use refraction index $n_2$ , as to the
index of refraction $n_1$ was found in the supposition, that this frequency
does not vary. Substituting (\ref{GL19}) in (\ref{GL27}), we get
the formula
\begin{eqnarray}
\label{GL28}
 \omega\;=\;\omega_0 n_2\;=\;\omega_0\left(1+\frac{\gamma M}{rc^2}\right),
 \end{eqnarray}
which coincides with the formula (\ref{GL25}). Thus in extended
space model for red shift is received the same expression, as in
general theory of relativity.

In papers \cite{OK11,OK12} Okun' has offered to consider the red
shift of a photon as change it of speed, momentum and wavelength,
but the frequency was assumed constant. He proceeded from a
dispersing ratio for a photon with zero mass in space with the
Schwarzchild metric
\begin{eqnarray}
\label{GL29}
g^{00}p_0p_0\;-\;g^{rr}p_rp_r\;=\;0.
\end{eqnarray}

The Schwarzchild metric is
\begin{eqnarray}
\label{GL30}
g^{00}\;=\;(1-\frac{r_g}r)^{-1},\;\;g^{rr}\;=\;(1-\frac{r_g}r),\\
\nonumber
r_g\;=\;\frac{2\gamma M}{c^2}.
\end{eqnarray}
Assuming, that $ñp_0=\hbar\omega=const$,  Okun' has received for relation of
a momentum $p_r$ from a radius $ r$ the expression
\begin{eqnarray}
\label{GL31}
p_r(r)\;=\;\frac{\hbar\omega}{c}(1-\frac{r_g}r)^{-1}\;=\;p_r(\infty)n_1,
\end{eqnarray}

Here $p_r(\infty)$ - is a momentum of the photon at infinity, where the
influence of gravitational field is absent.
Using connection between a momentum of a photon and its wavelength
$\lambda(r)$, we get the expression
\begin{eqnarray}
\label{GL32}
\lambda(r)=\frac{2\pi}{\omega}v=\frac{2\pi c}{\omega}(1-\frac{r_g}r)=
\frac {2\pi c}{\omega{n_1}}=\frac{\lambda(\infty)}{n_1}.
 \end{eqnarray}

For the speed of a photon $v(r)$  Okun' received the expression
\begin{eqnarray}
\label{GL33}
v(r)\;=\;\frac{\lambda(r)\omega}{2\pi}\;=\;c(1-\frac{r_g}r)\;=\;\frac c{n_1}.
\end{eqnarray}

Let's look now at transformation (\ref{GL26}) from the ESM point
of view. As the frequency of a photon remains constant, but vary
its momentum and mass the appropriate transformation must be
described by a rotation in the plane (XS) of the spaces G(1,4). As
the frequency does not vary, we take the refraction index $n_1$.
At such rotation the speed of a photon varies according to the
formula
\begin{eqnarray}
\label{GL34}
v\;=\;c\cos\psi\;=\;\frac cn_1.
 \end{eqnarray}

This formula coincides with the formula (\ref{AL18}) for
transformation of speed. Being repelled from it is possible to
receive the formula (\ref{GL32}), assigning change of a wavelength
of a photon, when photon hit in a gravitational field.

From a point of view of our model it is necessary to consider the
formula (\ref{GL26}) only as first approximation to an exact
result. Let's estimate correction appropriate to that in this
model the photon, hitting in area with $ n > 1$ gains a nonzero
mass. For this reason the part of photon energy can be connected
not to frequency, but with the mass. Let's estimate magnitude of
this energy for case, when photon frequency change, in case of
incident from a height H in a homogeneous gravitational field,
with acceleration of gravity g is measured. Such situation was
realized in well known Pound and Rebka experiments \cite{PR14}.
The energy change which appropriated to such frequency shift, is
equal
\begin{eqnarray}
\label{GL36}
\Delta E\;=\;\left(\frac{\hbar\omega}{c^2}\right)gH.
\end{eqnarray}
According to the formula (6) in the case of rotation in plane (TS) the photon
gains a mass
\begin{eqnarray}
\label{GL37}
m=\frac{\hbar\omega}{c^2}\sqrt{n^2-1}.
 \end{eqnarray}
The difference of potential energies in the point of emission and point of
absorption of a photon, which differ by height H, is equal
\begin{eqnarray}
\label{GL38}
\delta E\;=\;mgH\;=\;\left(\frac{\hbar\omega}{c^2}\right)gH\sqrt{n^2-1}.
\end{eqnarray}

Near to a surface of the Earth refraction index of gravitational
field is define by the formula (\ref{GL19}). Taking into
consideration an inequality (\ref{GL23}), one can get an
evaluation
\begin{eqnarray}
\label{GL39}
\delta E\;=\;mgH\;=\;\left(\frac{\hbar\omega}{c^2}\right)gH\sqrt{n^2-1}
\;\approx\;\\
\nonumber
 \left(\frac{\hbar\omega}{c^2}\right)gH\sqrt{\frac{2\gamma M}{Rc^2}}=\\
\nonumber
\left(\frac{\hbar\omega}{c^2}\right)gH\sqrt{\frac{2gR}{c^2}}\;\approx\;
\left(\frac{\hbar\omega}{c^2}\right)gH(2.5\cdot 10^{-5}).
\end{eqnarray}

We see, that correction to effect connected to emerging of the photons nonzero
mass, near to the Earth surface is only $10^{-5}$ from magnitude of the total
effect.

\bigskip
3) Delay of radar echo.

The appearance of radar echo delay is, that the time of light
distribution up to some object, and back, can differ in dependence
from that, does this light spread in a hollow, or in a
gravitational field. Such delay was measured in experiments on
location of Mercury and Venus \cite{SP15}. Such experiments give
satisfactory agreement with GR predictions. These experiments also
were analyzed in \cite{WB16}. Here we do not interesting to
analysis of these work. We want only to indicate that the
analytical expression for magnitude of delay of a radar echo in
ESM coincides what is received in GRT. This result can be obtained
from the fact that the photon time delay $\Delta t$ is calculated
from only from the photon velocity $v(t)$ \cite{OK11,OK12}. Let's
imagine that we locate Sun. In this case we have
\begin{eqnarray}
\label{GL40}
\Delta t=2\left(\int\limits_{R_s}^{r_e} \;\frac{dr}{v(r)}-
\int\limits_{R_s}^{r_e} \frac{dr}c\right).
 \end{eqnarray}

Here $R_s$ - is the radius of Sun, $r_g$ - is agravitational
radius of Sun, and $r_e$ - is a distance from Earth up to Sun. The
speed of light in a gravitational field is $v = \frac cn $.  As
here we deal with photons, as a refraction index it is necessary
to select $n = n_1$. Substituting it in (\ref{GL40}), we obtain
\begin{eqnarray}
\label{GL41}
\Delta t=2\frac{r_g}c\ln\frac{r_e}{R_s}.
 \end{eqnarray}

The formula (\ref{GL41}) coincides with expression for magnitude
of radar echo delay obtained in works \cite{OK12, WB16}.

\bigskip
4) Deviation angle of a light beam.

In general theory of relativity the magnitude of deviation angle
$\delta\psi$  of a light beam from a rectilinear trajectory in the
case of photon motion near to a massive body determine, deciding
the eyconal equation which defining trajectory of this beam in a
central-symmetrical gravitational field \cite{LL1}. In this case
one finds
\begin{eqnarray}
\label{GL42}
\delta\psi=4\frac{\gamma M}{Rc^2}.
 \end{eqnarray}
Here $M$ - is a mass of a body, and $R$ - is a distance at which
the light eam passes from a field center. As in this case speech
goes about photons motion it is necessary to select $n = n_1$.
Let's consider two beams - one passes precisely through an edge of
the Sun, and other at a distance x from it. It is supposed, that
 $$ h << R_s < r=\sqrt{x^2+R^2}.$$
  In the case of passing by these rays of a linear segment of length $dx$ the
  residual of optical paths will be
\begin{eqnarray}
\label{GL43}
\delta x=dxn_1(r)-dxn_1(r+h\cos\varphi)=\\
\nonumber
dx\left(1-\frac{r_g}r\right)-dx\left(1-\frac{r_g}{r+h\cos\varphi}\right)\
approx\\
\nonumber
\approx\;\frac{r_gh\cos\varphi}{r^2}dx .
\end{eqnarray}
To such difference of optical paths there corresponds an angle of a wave front
deviation
\begin{eqnarray}
\label{GL44}
\delta\varphi\approx\frac{\delta x}h\;=\;\frac{r_g\cos\varphi}{r^2}dx
\;=\;\frac{r_gR_s}{r^3}dx\;=\\
\nonumber
\frac{r_gR_sdx}{(x^2+R_s^2)^{3/2}}.
 \end{eqnarray}
Integrating this expression on $x$ from  $-\infty$ up to $+\infty$ we shall
receive deviation angle
\begin{eqnarray}
\label{GL45}
\varphi=r_g R_s\int\limits_{-\infty}^{\infty}\frac{dx}{(x^2+R_s^2)^{3/2}}
=2\frac{r_g}{R_s}=2\frac{\gamma M}{R_sc^2}.
 \end{eqnarray}
Expression (\ref{GL45}) yields half the angle (\ref{GL42}). This
epression is
 obtained in the geometrical-optics approximation ignoring the fact that
 according to the ESM the photon must acquire a mass in the gravitational
 field. For this effect we now estimate the second half caused by the fact
 that in the gravitational field the photon acquires a nonzero mass. In the
 case under discussion the value of this mass is of no importance, and we
 denote simply as $m_f$. We analyze the motion of a particle of mass $m_f$
 assuming this particle to move with impact parameter $R$ to the center of
 the gravitational field produced by the mass $M$. Let the motion of a
 particle in the (XY) plane be described by the Newton equation
\begin{eqnarray}
\label{GL46}
m_f\frac{d^2y}{dt^2} = -\gamma\frac{Mm_f}{r^2}\frac yr.
 \end{eqnarray}
Here $r^2=x^2+y^2$.

The photon mass $m_f$ is supposed to be constant. For this reason
we can exclude it from equation (\ref{GL46}). We assume that the
motion of the photon proceeds basically along the $X$ axis, and
the variable $y$ varies only slightly remaining close to the value
of the impact parameter $R$. We also consider the photon velocity
to remain constant all the time and to be equal to the velocity
$c$ of light in vacuum. Therefore, using the relationship
\begin{eqnarray}
\label{GL47}
y\;\approx\;R,\;\;x\;=\;ct,
\end{eqnarray}
we can transform equation (\ref{GL46}) to the form
\begin{eqnarray}
\label{GL48}
\frac{d^2y}{dx^2} = \gamma\frac{MR}{r^3}.
 \end{eqnarray}
After the first integration, we arrive at
\begin{eqnarray}
\label{GL49}
\frac{dy}{dx} = \frac{\gamma M}{c^2R}\frac x{\sqrt{x^2+y^2}}.
 \end{eqnarray}
Using this equation we can calculate the deflection angle
\begin{eqnarray}
\label{GL50}
\theta\;\approx\;\frac{dy}{dx}|_{-\infty}\;-\;\frac{dy}{dx}|_{\infty} =
2\frac{\gamma M}{c^2R}
 \end{eqnarray}
When the impact parameter $R$ equals the Sun radius $R_s$, the angle $\theta$
 coincides with the angle $\varphi$ given by the formula (\ref{GL45}). In the
 ESM these two effects are summed up to yield the total deflection angle
\begin{eqnarray}
\label{GL51}
\theta\;+\;\varphi\;=\;4\frac{\gamma M}{c^2R}.
 \end{eqnarray}
 The result (\ref{GL51}) coincides with formula (\ref{GL42}).

\bigskip
5) Perihelion precession of Mercury.

One more classical GR effect is the perihelion precession of
Mercury. It arises due to a space curvature the Newton's law of an
attraction is deformed. It reduces that the trajectory of a
particle becomes nonclosed, and after of each rotation it the
perihelion recessed at some angle. The magnitude of this rotation
is determined by the law of interaction of a central mass M and
mass m particles rotated around it. In case of a Schwarzchild
potential the force of interaction of these masses is \cite{ML17}
\begin{eqnarray}
\label{GL52}
F(r)=-\frac{\gamma Mm}{r^2\sqrt{1-R_g/r-v^2/c^2}}.
\end{eqnarray}
Here $R_g$ is the gravitational radius corresponding to the mass
$M$, and $v$ is the orbital velocity of a particle of mass $m$.
The velocity of Mercury orbital motion around the Sun is
approximately equal to 48 $km/s$, which r esults in the
relativistic correction $v^2/c^2\approx 5\times10^{-8}$. The
gravitational correction also attains $R_g/r\approx
5\times10^{-8}$ and is very close to the relativistic one. One can
assume that the particle mass $m$ in formula (\ref{GL52}) depends
on both the distance and velocity. However as both these
corrections are small one can write the total transformation of
mass $m$ in the form
\begin{eqnarray}
\label{GL53}
m\;\rightarrow \;m\left(1+\frac{\gamma M}{rc^2}+\frac12\frac{v^2}{c^2}\right).
\end{eqnarray}
The calculation using the expression for the force (\ref{GL52}) with allowance
for approximation (\ref{GL53}) yields the Mercury perigee shift to the
observed one.

Let's now analyze this problem from the stand point of ESM. We deal with a
nonzero mass particle placed into the gravitational field. Since one cannot
already ignore the relativistic corrections, we will use the refraction index
in the form (\ref{GL21}). In this case the particle gets from a domain with a
refraction index $n=1$ to a domain with refraction index $n'$ because of the
variation of the force acting on it, i.e. because of a change of its energy.
Therefore the (TS) rotation in $G(1,4)$ space should be used. In the case of
this rotation the massive vector is transformed in accordance with formula
(\ref{AL517}) and a massive particle changes its mass obeying the formula
(\ref{AL518}). Under such transformation particles of masses $m_+$ and $m_-$
can arise from of mass $m$.
\begin{eqnarray}
\label{GL54}
m_+ = me^{\theta_+}=m(n+\sqrt{n^2-1}),\\
\nonumber
m_- = me^{\theta_-}=m(n-\sqrt{n^2-1}).
\end{eqnarray}
We assume that a macroscopic massive body has an equal number of particles
transformed according to laws (\ref{GL54}). We will us for this situation the
average transformation law
\begin{eqnarray}
\label{GL55}
m \rightarrow mn'_2=m\left(1+\frac{\gamma M}{rc^2}+\frac12\frac{v^2}{c^2}
\right).
\end{eqnarray}
As one can see formulas (\ref{GL53}) and (\ref{GL55}) coincide.

\section{Radioastron mission - visible size of bubble objects}

In resent years new phenomena which go beyond the tradional GRT-based concepts
concerning the structure of Universe, have been discovered by astronomers.
The essence of these phenomena is basically as follows:

1) The bulk of the Universe mass (more than 90\%) is dark matter and hidden
energy which is associated with the cosmic vacuum. 2) This dark matter does
not emit electromagnetic radiation and does not interact with it, but shows
gravitational properties. 3) The cosmic vacuum possesses negative pressure or,
in other words shows antigravitational properties, which determine the current
dynamics of Universe expansion. 4) Usual massive objects are surrounded by
dark-matter halo.

The ESM gives us an approach for the explanation of these
phenomena. As was allready shown the motion in the additionaL
fifth dimension corresponds to change of the particle rest mass.
When the photon gets into the external field, it acquires a
nonzero mass which can be either positive or negative. We suppose
that the inertia of such mass is always positive, and it is only
its gravitational properties that can have different signes. In a
pair of photons born in an external field one has a positive and
the other a negative mass. According to ESM the dark matter
consists of massive photons. Positive-mass photons are
concentrated around massive stars and black holes and form their
halos. Negative-mass photons are throws away into the free cosmic
space where they create an antigravitating vacuum with negative
pressure. Hence,in our opinion, the dark matter mostly consists of
positive-mass photons and the dark energy is generated by
negative-mass photons. Different ways of assigning a nonzero mass
to a photon are discussed in review  \cite{RV4}. The possibility
of the existence of bodies with a negative mass was discussed in
\cite{BN18}. Attension has recently been drawn by a new
gravitational model, the so-called gravstar, or gravitational
condensate star  \cite{MM19}. It has been proposed as an
alternative to black holes. These objects correspond to the
solutions of Einstein equations which outside a region occupied by
masses coincides with the Schwarzschild solution. Inside it there
is another nonsingular solution, and so the metric as a whole
appeares to be nonsingular. The gravstar structure is similar to
that of a bubble. A bubble has a rigid dense shell which is
stressed because of a liquid substance pressing out from inside.
This particular model is now typically used to explain the nature
of some observed objects. It is shown that in ESM model formation
of bubble gravitational structures is possible. In the frame of
ESM one can obtain the follow physical picture. B ubble
gravitational objects have a halo formed by dark matter generated
by photons with a positive mass. Now it becomes possible to
predict some future results of visible size of supermassive
objects in our Universe due to new stage of experimental astronomy
development in the RadioAstron Project \cite{RA20,RA21,RA22}. As
to in RadioAstron Project has to reached unprecedented angular
resolution equal 0,00001" it is becomes possible to distinguish
real size of such objects active galactic nuclei to investigate
the size of supermassive objects applying for black hole role and
to obtain direct results of comparison event horizon with its
radius. In the case when the visible radius of supermassive object
will not exceed the gravitational radius we will agree that this
object is real black hole. In the case when visible radius will be
in 2 to 3 times larger than gravitational radius appropriated to
the mass of supermassive object - we will discuss the nature of
such objects with taking into account that one of the possible
answers is the gravastar.

\section{ V838 Monocerotis explosive outburst - local Big Bang or light echo?}
The ESM can be applied to analysis of new phenomenon taking place
in the Universe. They did not have until now any generally
accepted physical explanation. One of these mysterious phenomena
is an explosion of the star V838 Mon. In 2002, the previously
unknown variable star V838 Monocerotis brightened suddenly by a
factor of about $10^4$. Unlike a supernova or nova, V838 Mon did
not explosively eject its outer layers; rather, it simply expanded
to become a cool supergiant with a moderate-velocity stellar wind.
Investigation of this phenomenon show, that hen combined with the
high luminosity and unusual outburst behavior, these
characteristics indicate that V838 Mon represents a hitherto
unknown type of stellar outburst, for which we have no completely
satisfactory physical explanation  \cite{ST23}. The morphing
sequence of six images taken by Hubble's Advanced Camera of V838
explosion could be found at \cite{ST24}. V838 image evolution was
attributed not to a cloud expanding (as are normal supernovas) but
light echoing (hitting different parts) of an interstellar cloud.
A set of mechanism were proposed to explain outburst see reference
in \cite{ST25}. There is some discussion with the nature of light
echo and source of light echo materials as to due to estimation
\cite{ST26} mass light echo materials is about 90-150 mass of the
Sun. In the frame of ESM one can consider a possibility to explain
V838 explosive outburst \cite{TA9,TA10}.  We could interpret this
phenomenon as local Big Bang. Thus we try to consider V838
evolution in ESM formalism as a new space local birth. In our
(4+1)-dimensional model V838 expansion could be explained as
evolution of high-dencity complex field consisting from 4 fields -
scalar field Q, gravitational field G, electic feield E and
magnetic field H. We also discuss the nature of origin SiO maser
emission from the direction of V838 \cite{ST25}. In the framework
of our ESM model we also explain the absence of large molecular
lines shift in the spectrum of CO and AlO \cite{ST25,ST26}.

\section{Conclusion}
In given work the generalization of Einstein's Special theory of
relativity is proposed. It is  the (4+1)-dimensional Extended
space model. In this model gravity and electromagnetism are
unified into a single field. The gravitational effects such as the
speed of escape, gravitational red shift and deflection of light
can be found algebraically by the rotations in the (1+4)
dimensional space. The dark matter and hidden energy get natural
interpretation in the frame of this model. Thus, in ESM frameworks
there is a mechanism, according to which in a neighborhood of
massive gravitational objects the halo consisting of photons with
positive masses can be formed. These halos we associate with a
dark matter.  The photons with negative masses are concentrated
far from massive objects. These photons will create areas of a
dark energy. Such areas are characterized by negative pressure and
exhibits properties of antigravitation.

 We also consider a possibility to explain V838 explosive outburst with
 its
 help. We suppose that this  phenomenon can be  interpreted as a local
 Big Bang. In this case we see the movement of thr space itself, and the rate
 of its expansion is seen as the rate of expansion of stars shell.
 The observed rate of expansion may exceed the speed of light.



\begin{thebibliography}{19}
\bibitem{LL1} L.D. Landau  and  E.M.Lifshitz, {\it The classical theory of
fields }
  (World Scientific, Singapore,  1994).

\bibitem{SW2} S.S. Schweber, {\it An Introduction to Relativistic Quantum Fiel
Theory} (World Scientific, Singapore, 1961).


\bibitem{GN3} V.L. Ginzburg, {\it Theoretical Physics and Astrophysics}
(NAUKA, Moscow, 1981).

\bibitem{RV4} L.A. Rivlin,  {\em Physics-Uspekhi}  {\bf 40}, 291 (1997).

\bibitem{OK5} L.B. Okun',  {\em Physics-Uspekhi}  {\bf 32}, 629 (1989).

\bibitem{KN6} S. Kobayashi and K. Nomizu, {\it Foundation of Differential
Geometry. Vol. 1}
( NEW YORK LONDON,1963).

\bibitem{TA7} D.Yu. Tsipenyuk  and V.A. Andreev, {\em Bulletin of the Lebedev
Physics Institute}
 {\bf 6}, 23 (2000); arXiv:gr-qc/0106093, 2001.

\bibitem{TA8} D.Yu. Tsipenyuk  and V.A. Andreev, {\em Bulletin of the Lebedev
Physics Institute}
 {\bf 6}, 1 (2002); arXiv:physics/0302006, 2003.

\bibitem{TA9} D.Yu. Tsipenyuk  and V.A. Andreev, {\it \em Bulletin of the
Lebedev Physics Institute},
 {\bf 9}, 10 (2004); arXiv:physics/0506002, 2004.

\bibitem{TA10} D.Yu. Tsipenyuk  and V.A. Andreev, {\em Bulletin of the
Lebedev Physics Institute}
 {\bf 10}, 13 (2004);  arXiv:physics/0407144, 2004.

 \bibitem{OK11} L.B. Okun', K.G. Selivanov, and V.L. Telegdi,
 {\em Physics-Uspekhi}  {\bf 42}, 1141 (1999).

\bibitem{OK12} L.B. Okun',  {\em Physics-Uspekhi}  {\bf 43}, 1366 (2000).

\bibitem{CL13} R.L. Collins, {\it Gravity slows the speed of light},
APS reprint server, (8/9/97).
  (World Scientific, Singapore,  1994).

\bibitem{PR14} R.V.Pound, and G.A. Rebka, {\em Phys. Rev. Lett.}  {\bf 4},
337 (1960).

\bibitem{SP15} I.I. Shapiro, {\em Phys. Rev. Lett.}  {\bf 13}, 789 (1964).


\bibitem{WB16} S. Weinberg, {\it Gravitation and Cosmology}
(John Willey and Sons, New York,  1972).

\bibitem{ML17} C. Moeller, {\it The Theory of Relativity}   (
Oxford: Clarendon Press,  1972).

\bibitem{BN18} H. Bondi, {\em Rev. Mod. Phys.}  {\bf 29}, 423 (1957).


 \bibitem{MM19}  P.O. Mazur and E. Mottola, {\it Gravitational Condensate
 Stars: An Alternative to
Black Holes}, arXiv:gr-qc/0109035.


\bibitem{RA20} RadioAstron Project
   http://www.asc.rssi.ru/radioastron/index.htm.

\bibitem{RA21} RADIOASTRON The Ground -Space Interferometer:http://
www.asc.rssi.ru/radioastron/files=bookleten:pdf.

\bibitem{RA22}  Radioastron Handbook http://www.asc.rssi.ru
/radioastron/documents/rauh/en/rauh.pdf.


 \bibitem{ST23}  H. E. Bond, A, Henden, Z. G. Levay, et al.,
 {\it An energetic stellar outburst accompanied by circumstellar light
 echoes},   arXiv/astro-ph/0303513.

\bibitem{ST24}  V838 Mon outburst movie http://www.spacetelescope.org/videos
/heic0503a/.

 \bibitem{ST25}   T. Kamiñski, {\it Extended CO emission in the field of the
 light echo of V838 Mon}, arXiv:0801.1689v1.

 \bibitem{ST26}   T. Kamiñski, R. Tylenda, S. Deguchi, {\it A molecular cloud
 within the light echo of V838 Monocerotis}, arXiv:1102.5237v1




\end{thebibliography}
\end{document}